\documentclass[10pt,draft]{dis03}
\usepackage{epsf,amsmath}

\def\beq{\begin{equation}}
\def\eeq{\end{equation}}
\def\beqa{\begin{eqnarray}}
\def\eeqa{\end{eqnarray}}
\def\ifm{\ifmmode}
\def\msb{\ifm \overline{\rm MS}\,\, \else $\overline{\rm MS}\,\, $\fi}
\def\msbns{\ifm \overline{\rm MS}\, \else $\overline{\rm MS}\, $\fi}
\def\e{\epsilon}
\def\a{\alpha}

\begin{document}

\title{EXPONENTIATION AT PARTONIC THRESHOLD \\
FOR THE DRELL-YAN CROSS SECTION}

\author{T.O.~Eynck$^a$, E.~Laenen$^{a,b}$ and L.~Magnea$^c$\\
$^a$ NIKHEF Theory Group\\
Kruislaan 409, 1098 SJ Amsterdam, 
The Netherlands\\
$^b$ Institute for Theoretical Physics, Utrecht University\\
Leuvelaan 4, 3584 CE Utrecht, The Netherlands\\
$^c$ Dipartimento di Fisica Teorica, Universit\`a di Torino\\
and INFN, Sezione di Torino\\
Via P. Giuria 1, I-10125, Torino, Italy}

\maketitle

\begin{abstract}
\noindent The techniques leading to the resummation of threshold
logarithms in the Drell-Yan cross section and other processes can be
used to show that also terms independent on the Mellin variable $N$
exponentiate. Comparison with explicit two-loop calculations shows
that within this class of terms the exponentiation of the one-loop
result together with the running of the coupling is the dominant
effect at two loops.
\end{abstract}

\section{Introduction}

Threshold resummations~\cite{Sterman:1987aj} have been an active field
of study in QCD for over two decades. They have a direct relevance to
phenomenology~\cite{Giele:2002hx}, since they extend the range of
applicability of perturbative methods for the computation of hard
cross sections, but they are also interesting from a purely
theoretical point of view, since they provide a method to probe the
interface of perturbative and nonperturbative physics.

The central issue which is addressed by threshold resummations is the
physics of inhibited radiation. At high energy, in a production
process which forces the radiation of gauge bosons in the final state
to be either soft or collinear to the hard partons carrying the bulk
of the invariant mass, the validity of the perturbative expansion is
jeopardized by the presence of large double (Sudakov) logarithms to
all orders. The logarithms are the remainders of soft and collinear
singularities, after the cancellations required by the mandatory
infrared safety of the observable; they are large, because such a
process has two disparate scales: the hard scale $s$ and the scale of
soft radiation $\tau s$ ($\tau \ll 1$); they can be exponentiated and
resummed, thanks to the factorizability and universality properties of
soft and collinear radiation.

To account for the constraint imposed by momentum conservation, the
exponentiation of soft contributions takes place after a Mellin (or
Laplace) transform with respect to the variable vanishing at
threshold, $\tau$.  In terms of the Mellin variable $N$, conjugate to
$\tau$, one finds a hierarchy of contributions, with an increasingly
mild behavior at large $N$, corresponding to a decreasing level of
infrared sensitivity at small $\tau$. Sudakov logarithms are the only
terms growing with $N$, as $\alpha_s^n \log^k N$, with $k \leq 2
n$. Then one finds terms independent of $N$, then terms suppressed by
powers of $N$, which however may still have logarithmic behavior in
$N$, such as $\alpha_s^n (\log^k N)/N$, with $k \leq 2 n - 1$.

Most of the work done in the past on the subject of threshold
resummations has focused on Sudakov logarithms, which have been shown
to exponentiate to all logarithmic accuracies, and are currently being
explicitly evaluated at NNL level~\cite{Vogt:2000ci}. It was clear
from the beginning~\cite{Parisi:1980xd}, however, that at least some
of the $N$-independent terms arising in the relatively simple case of
the Drell-Yan process were both numerically important and
exponentiating together with the logarithms. Furthermore, at least a
subset of the logarithmic terms suppressed by a power of $N$ can also
be resummed by similar techniques, and it has been shown that these
terms have a considerable impact on cross sections of phenomenological
interest~\cite{Kramer:1996iq}. It is therefore of both practical and
theoretical interest to study the possibility of extending currently
available resummation techniques beyond the case of Sudakov
logs. Here, we will briefly discuss the exponentiation of
$N$-independent terms in the relatively simple cases of the DIS
structure function $F_2$ and of the Drell-Yan cross section, following
\cite{Eynck:2003fn}.

\section{Factorization and exponentiation for $N$-independent terms}

The resummation of threshold logarithms can be derived by factorizing
the relevant cross section to isolate soft and collinear enhancements.
For the partonic Drell-Yan cross section this factorization takes the
form~\cite{Sterman:1987aj}
\beqa
 \label{om}
 \omega (N,\e) = |H_{\mathrm{DY}}|^2 \, \psi(N, \e)^2 
 \, U(N) + {\cal O}(1/N) \, ,
\eeqa
where $\psi(N, \e)$ is a parton distribution containing singular
collinear contributions, while $U(N)$ is an eikonal cross section
responsible for coherent soft radiation.  The DIS structure function
$F_2$ obeys a similar factorization
\beq
\label{F2}
 F_2 (N, \e) = |H_{\mathrm{DIS}}|^2 \, \chi(N, \e) 
 \, V(N) \, J(N) + {\cal O}(1/N) \, ,
\eeq
where $\chi$ and $V$ play the same role as $\psi$ and $U$,
respectively, while $J(N)$ is a jet function summarizing the effect of
collinear final state radiation.

These factorizations are valid up to corrections suppressed by powers
of $N$, and thus they include $N$-independent terms. The key
observation is that it is possible to separate, within each function,
real emission diagrams from purely virtual contributions, and then to
express all virtual (an thus $N$-independent) functions in terms of
the quark form factor. Specifically one finds that
\beqa
 \omega (N, \e) & = & |\Gamma(Q^2,\e)|^2 \, \psi_R(N, \e)^2 \,
 U_R(N, \e) + {\cal O}(1/N)~, \label{arifac1} \\
 F_2 (N, \e)  & = &  |\Gamma(- Q^2, \e)|^2 \, \chi_R(N, 
 \e) \, V_R(N, \e) \, J_R(N,\e) + {\cal O}(1/N) \,.
\label{arifac2}
\eeqa
Each factor in Eqs.~(\ref{arifac1}) and (\ref{arifac2}) is now a pure
exponential: for real emission contributions this was established
in Ref.~\cite{Sterman:1987aj}, while for the form factor in
dimensional regularization it was proven in
Ref.~\cite{Magnea:1990zb}. Taking the ratio of (\ref{arifac1}) and
(\ref{arifac2}) one finds an exponentiated expression for the
factorized DIS-scheme Drell-Yan cross section, valid up to corrections
suppressed by powers of $N$, which can be organized in the form
\beqa
\label{findis}
 &&  \hspace{-10pt} \widehat{\omega}_{\mathrm{DIS}} (N) =  
 \left|\frac{\Gamma(Q^2, \e)}{\Gamma(-Q^2, \e)} \right|^2
 \exp \Big[ F_{\mathrm{DIS}}(\a_s) \Big] ~\exp \Bigg[ \int_0^1 \! dz \,
 \frac{z^{N - 1} - 1}{1 - z} \\ 
 && \hspace{-24pt} \Bigg\{ 2 \int_{(1 - z)Q^2}^{(1 - 
 z)^2 Q^2} \! \frac{d \xi^2}{\xi^2} \, A \left( \a_s( \xi^2) \right) 
 - 2 B \left(\a_s \left((1 - z) Q^2 \right) \right) + 
 D \left( \a_s \left((1 - z)^2 Q^2 \right) \right) 
 \Bigg\} \Bigg] \nonumber .
\eeqa
Explicit expressions for the various functions involved are listed 
in~\cite{Eynck:2003fn}.

It is not difficult to generalize this result to the \msb scheme. The
Mellin transform of the \msb quark distribution, $\phi(N, \e)$, can,
in fact, be written in exponential form, up to corrections suppressed
by powers of $N$. Further, as shown in~\cite{Eynck:2003fn}, it is
possible to factorize real and virtual contributions to $\phi(N, \e)$,
so that virtual poles precisely cancel those arising from the quark
form factor. The factorized \msbns-scheme Drell-Yan cross section can
then be written as the product of two finite exponential factors, one
associated with real gluon emission, and one with purely virtual
graphs. The final result can be organized in the form
\beqa
 && \hspace{-17pt} \widehat{\omega}_{\msb}(N) = \left|\frac{\Gamma(Q^2,
 \e)}{\Gamma(- Q^2, \e)} \right|^2
 \left( \frac{\Gamma(-Q^2,\e)}{ \phi_V(Q^2,\e)} \right)^2
 \exp \Big[ F_{\msbns} (\a_s) \Big] ~\exp \Bigg[ \int_0^1 \! dz \,
 \frac{z^{N - 1} - 1}{1 - z} 
 \nonumber \\ &&
 \Bigg\{ 2 \, \int_{Q^2}^{(1 - z)^2 Q^2} 
 \frac{d \mu^2}{\mu^2} \, A \left(\a_s(\mu^2) \right)
 + D \left(\a_s \left((1 - z)^2 Q^2 \right) \right) \Bigg\} 
 \Bigg]~.
\label{finms}
\eeqa
Again, explicit expressions for the functions involved are given in
Ref.~\cite{Eynck:2003fn}. Eqs.~(\ref{findis}) and (\ref{finms}) are
our main results: they generalize currently available resummations for
the Drell-Yan partonic cross section to include the exponentiation of
all $N$-independent terms. Clearly, the same reasoning and a similar
formula apply for the partonic DIS structure function $F_2$ factorized
in the \msb scheme.

\section{Usage and impact of the exponentiation}

It should be clear that the exponentiation of $N$-independent terms
does not have the same predictive power as the resummation of Sudakov
logarithms. In fact, for example, a one-loop calculation suffices to
determine exactly the coefficients of the leading Sudakov logs to all
orders, whereas constant terms, although exponentiating, receive
nontrivial contributions order by order. The resummation formulas in
Eqs.~(\ref{findis}) and (\ref{finms}) can however be used in practice
in at least two different ways.

First of all, for the purpose of analytic calculations, on can make
use of the fact that all functions appearing in the factorizations
(\ref{om}) and (\ref{F2}) have precise operator definitions, as well
as of the fact that in the resummations (\ref{findis}) and
(\ref{finms}) real and virtual contributions are separately
finite. These two facts provide alternative, and often simpler,
methods to perform and test analytic calculations at the resummed
level, without having to compute the full cross section at the
relevant perturbative order. As an example, in
Ref.~\cite{Eynck:2003fn} we were able to compute the second order
perturbative coefficient of the function $D$, using only information
arising from virtual graphs, and reproducing the result previously
obtained~\cite{Vogt:2000ci,Magnea:1991qg} by matching to the exact
two-loop cross section.

From the point of view of phenomenology, although exponentiation of
the one-loop result does not suffice to determine exactly any specific
perturbative coefficient at higher orders, it remains true that a
nontrivial subset of higher order corrections originate from
exponentiation. This subset provides a well-founded estimate of the
uncertainty of the perturbative calculation due to unknown higher
order corrections. To test this fact, in Ref.~\cite{Eynck:2003fn}, we
estimated the numerical values of $N$-independent terms at two loops,
by using only one-loop information in the exponent, together with
renormalization group running. We find that exponentiation predicts
three quarters of the exact answer for these terms in the DIS scheme,
whereas for the \msb scheme the prediction exceeds the exact value by
about $70\%$. In both cases, we conclude that exponentiation gives a
fairly reliable estimate of the complete calculation.

We conclude by noting that these results may be seen as a first step
towards further generalizations of soft gluon resummation: first of
all, it would be interesting to study the exponentiation of
$N$-independent terms for more general QCD cross sections, in the
presence of nontrivial color exchange; furthermore, a precise
formulation of the exponentiation of Sudakov logarithms suppressed by
a power of $N$ would be of considerable phenomenological interest.

\end{document}